\documentclass[aoas,MSNbibl,nameyear,seceqn,dvips]{arximspdf}
\usepackage{graphicx}

\doi{10.1214/10-AOAS410}
\volume{5}
\issue{2A}
\pubyear{2011}
\firstpage{1057}
\lastpage{1080}

\makeatletter

\newtheorem{proposition}{Proposition}
\makeatother

\begin{document}
\begin{frontmatter}

\title{Robust graphical modeling of gene networks using classical and alternative
$\bolds{t}$-distributions\thanksref{T1}}
\runtitle{Robust graphical modeling using $t$-distributions}

\begin{aug}
\author[A]{\fnms{Michael} \snm{Finegold}\corref{}\ead[label=e1]{finegold@uchicago.edu}}
\and
\author[A]{\fnms{Mathias} \snm{Drton}\thanksref{t2}\ead[label=e2]{drton@uchicago.edu}}
\thankstext{T1}{Supported by
NSF Grant DMS-07-46265.}
\thankstext{t2}{Supported by an Alfred P. Sloan Fellowship.}
\runauthor{M. Finegold and M. Drton}
\affiliation{University of Chicago}
\address[A]{Department of Statistics\\
University of Chicago\\
5734 S. University Ave.\\
Chicago, Illinois  60637\\
USA\\
\printead{e1}\\
\phantom{E-mail: }\printead*{e2}} 
\end{aug}

\received{\smonth{12} \syear{2009}}
\revised{\smonth{7} \syear{2010}}

\begin{abstract}
Graphical Gaussian models have proven to be useful tools for exploring
  network structures based on multivariate data.  Applications to studies
  of gene expression have generated substantial interest in these models,
  and resulting recent progress includes the development of fitting
  methodology involving penalization of the likelihood function.  In this
  paper we advocate the use of multivariate $t$-distributions for more
  robust inference of graphs.  In particular, we demonstrate that penalized
  likelihood inference combined with an application of the EM algorithm
  provides a computationally efficient approach to model selection in the
  $t$-distribution case.  We consider two versions of multivariate $t$-distributions,
   one of which requires the use of approximation techniques.
  For this distribution, we describe a Markov chain Monte Carlo EM
  algorithm based on a Gibbs sampler as well as a simple variational
  approximation that makes the resulting method feasible in large problems.
\end{abstract}


\end{frontmatter}

\section{Introduction}\label{sec:introduction}

Graphical Gaussian models have attracted a lot of recent interest.  In
these models an observed random vector $Y=(Y_1,\dots,Y_p)$ is assumed
to follow a multivariate normal distribution
$\mathcal{N}_p(\mu,\Sigma)$, where~$\mu$ is the mean vector and
$\Sigma$ the positive definite covariance matrix.  Each model is
associated with an undirected graph $G=(V,E)$ with vertex set
$V=\{1,\dots,p\}$, and defined by requiring that for each nonedge
$(j,k)\notin E$, the variables $Y_j$ and $Y_k$ are conditionally
independent given all the remaining variables $Y_{\setminus\{j,k\}}$.
Here, $\setminus\{j,k\}$ denotes the complement $V\setminus\{j,k\}$.
Such pairwise conditional independence holds if and only if
$\Sigma_{jk}^{-1}=0$; see \cite{lauritzen} for this fact and general
background on graphical models.  Therefore, inferring the graph
corresponds to inferring the nonzero elements of $\Sigma^{-1}$.

Classical solutions to the model selection problem include
constraint-ba\-sed approaches that test the model-defining conditional
independence constraints, and score-based searches that optimize a
model score over a~set of graphs.  A review of this work can be found
in \cite{drtonperlman}. Recently, however, penalized likelihood
approaches based on the one-norm~of the concentration matrix
$\Sigma^{-1}$ have become increasingly popular. Meinshau\-sen and B\"uhlmann \citeyear{meinshausen}
proposed a method that uses lasso regressions of each variable $Y_j$ on
the remaining variables $Y_{\setminus
  j}:=Y_{\setminus\{j\}}$.  In subsequent work, \cite{yuan} and \cite{banerjee}
discuss the computation of the exact solution to the convex
optimization problem arising from the likelihood penalization. Finally,
\cite{friedmanhastie} developed the \textit{graphical lasso (glasso)},
which is a computationally efficient algorithm that maximizes the
penalized log-likelihood function through coordinate-descent.  The
theory that accompanies these algorithmic developments supplies
high-dimensional consistency properties under assumptions of graph
sparsity; see, for example, Ravikumar et
al.~(\citeyear{ravikumar}).\looseness=-1

Inference of a graph can be significantly impacted, however, by
deviations from normality.  In particular, contamination of a handful
of variables in a~few experiments can lead to a drastically wrong
graph.  Applied work thus often proceeds by identifying and removing
such experiments before data analysis, but such outlier screening can
become difficult with large data sets.  More importantly, removing
entire experiments as outliers may discard useful information from the
uncontaminated variables they may contain.

The existing literature on robust inference in graphical models is
fairly limited.  One line of work concerns constraint-based approaches
and adopts robustified statistical tests [\cite{kalisch}].  An approach
for fitting the model associated with a given graph using a robustified
likelihood function is described in \cite{miyamura}.  In some cases
simple transformations of the data may be effective at minimizing the
effect of outliers or contaminated data on a~small scale.  A normal
quantile transformation, in particular, appears to be effective in many
cases.

In this paper we extend the scope of robust inference by providing a
tool for robust model selection that can be applied with highly
multivariate data.  We build upon the \textit{glasso} of
\cite{friedmanhastie}, but model the data using multivariate
$t$-distributions.  Using the EM algorithm, the \textit{tlasso} methods
we propose are only slightly less computationally efficient than the
\textit{glasso} but cope rather well with contaminated data.

The paper is organized as follows.  In Section \ref{sec:glasso} we
review maximization of the penalized Gaussian log-likelihood function
using the \textit{glasso}.  In Section~\ref{sec:t-dist} we introduce
the classical multivariate $t$-distribution and describe maximization
of the (unpenalized) log-likelihood using the EM algorithm.  In
Sec\-tion~\ref{sec:tlasso} we combine the two techniques into the \textit{tlasso}
to maximize the penalized log-likelihood in the multivariate $t$ case.
In Section \ref{sec:alasso} we introduce an\vadjust{\eject} alternative multivariate
$t$-distribution and describe how inference can be done using
stochastic and variational EM.  In Section \ref{sec:simulations} we
compare the \textit{glasso} to our $t$-based methods on simulated data.  Finally, in
Section~\ref{sec:gene-expression-data} we analyze two different gene expression
data sets using the competing methods.  Our findings are summarized in
Section \ref{sec:discussion}.\


\section{Graphical Gaussian models and the graphical lasso}\label{sec:glasso}

Suppose we observe a sample of $n$ independent random vectors
$Y_1,\dots, Y_n\in\mathbb{R}^p$ that are distributed according to the
multivariate normal distribution $\mathcal{N}_p(\mu,\Sigma)$.
Likelihood inference about the covariance matrix $\Sigma$ is based on
the log-likelihood function\vspace*{-2pt}
\[
\ell(\Sigma) = -\frac{np}{2}\log (2\pi)-\frac{n}{2}\log\det(\Sigma) -
\frac{n}{2} \operatorname{tr}(S\Sigma^{-1}),\vspace*{-2pt}
\]
where the empirical covariance matrix\vspace*{-2pt}
\[
S = (s_{jk})= \frac{1}{n} \sum_{i=1}^n (Y_i-\bar Y)(Y_i-\bar Y)^T\vspace*{-2pt}
\]
is defined based on deviations from the sample mean $\bar Y$.  Let
$\Theta = (\theta_{jk})= \Sigma^{-1}$ denote the ($p\times
p$)-concentration matrix. In penalized likelihood methods a~one-norm
penalty is added to the log-likelihood function, which effectively
performs model selection because the resulting estimates of $\Theta$
may have entries that are exactly zero. Omitting irrelevant factors and
constants, we are led to the problem of maximizing the function
\begin{equation}  \label{eq:pen-lik}
\log \det(\Theta) - \operatorname{tr}(S\Theta) - \rho \|\Theta\|_1
\end{equation}
over the cone of positive definite matrices, where $\|\Theta\|_1$ is
the sum of the absolute values of the entries of $\Theta$. The
multiplier $\rho$ is a positive tuning parameter. Larger values of
$\rho$ lead to more entries of $\Theta$ being estimated as zero.
Cross-validation or information criteria can be used to tune $\rho$.

The \textit{glasso} is an iterative method for solving the convex
optimization problem with the objective function in (\ref{eq:pen-lik}).
Its updates operate on the covariance matrix $\Sigma$. In each step one
row (and column) of the symmetric matrix $\Sigma$ is updated based on a
partial maximization of (\ref{eq:pen-lik}) in which all but the
considered row (and column) of $\Theta$ are held fixed.  This partial
maximization is solved via coordinate-descent as briefly reviewed next.

Partition off the last row and column of $\Sigma=(\sigma_{jk})$ and $S$
as
\[
\Sigma =
\pmatrix{
  \Sigma_{\setminus p,\setminus p} & \Sigma_{\setminus p,p} \vspace*{1pt} \cr
  \Sigma_{\setminus p,p}^T & \sigma_{pp}},\qquad
S= \pmatrix{
    S_{\setminus p,\setminus p} & S_{\setminus p,p}\vspace*{1pt}  \cr
    S_{\setminus p,p}^T & s_{pp}}.
\]
Then, as shown in \cite{banerjee}, partially maximizing
$\Sigma_{\setminus p,p}$ with $\Sigma_{\setminus p,\setminus p}$ held
fixed yields $\Sigma_{\setminus p,p}=\Sigma_{\setminus p,\setminus p}
\beta^*$, where $\beta^*$ minimizes
\[
\|(\Sigma_{\setminus p,\setminus p})^{1/2} \beta - (\Sigma_{\setminus
p,\setminus p})^{-1/2} S_{\setminus p,p} \|^2 + \rho \|\beta\|_1
\]
with respect to $\beta\in\mathbb{R}^{p-1}$.  The \textit{glasso} finds
$\beta^*$ by coordinate descent in each\vadjust{\eject} of the coordinates
$j=1,\dots,p-1$, using the updates
\[
\beta^*_j = \frac{T (s_{jp}-\sum_{k<p,k \ne j} \sigma_{kj}
      \beta^*_k, \rho )}{\sigma_{jj}},
\]
where $T(x,t)=\operatorname{sgn}(x)(|x|-t)_+$. The algorithm then cycles
through the rows and columns of $\Sigma$ and $S$ until convergence.
The diagonal elements are simply $\sigma_{pp}=s_{pp} + \rho$.  See
\cite{friedmanhastie} for more details on the method.


\section{Graphical models based on the $t$-distribution}\label{sec:t-dist}
\subsection{Classical multivariate $t$-distribution}

The classical multivariate $t$-dis\-tribution $t_{p,\nu}(\mu,\Psi)$ on
$\mathbb{R}^p$ has Lebesgue density
\begin{equation}  \label{eq:tdensity}
  f_\nu(y;\mu,\Psi) = \frac{\Gamma ((\nu+p)/2) |\Psi|^{-1/2}}{(\pi
  \nu )^{p/2} \Gamma
  (\nu/2) [1 + \delta_y (\mu, \Psi)/\nu]^{(\nu+p)/2}}
\end{equation}
with $\delta_y(\mu,\Psi) = (y-\mu)^T \Psi^{-1} (y-\mu)$ and
$y\in\mathbb{R}^p$.  The vector $\mu\in\mathbb{R}^p$ and the positive
definite matrix $\Psi=(\psi_{jk})$ determine the first two moments of
the distribution.  If $Y\sim t_{p,\nu}(\mu,\Psi)$ with $\nu > 2$
degrees of freedom, then the expectation is $\mathbb{E}[Y]=\mu$ and the
covariance matrix is $\mathbb{V}[Y]=\nu/(\nu-2)\cdot\Psi$.  From here
on we will always assume $\nu>2$ for the covariance matrix to exist.
For notational convenience and to illustrate the parallels with the
Gaussian model, we define $\Theta= (\theta_{jk})=\Psi^{-1}$.

\begin{figure}[b]

\includegraphics{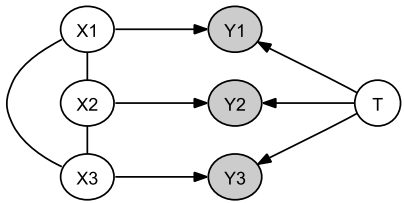}

\caption{Graph representing the process generating a multivariate
  $t$-random vector $Y$ from a latent Gaussian random vector $X$ and a
  single latent Gamma-divisor.}\label{fig:standard}
\end{figure}

If $X\sim\mathcal{N}_p(0,\Psi)$ is a multivariate normal random vector
independent of the Gamma-random variable $\tau\sim\Gamma(\nu/2,\nu/2)$,
then $Y=\mu+X/\sqrt{\tau}$ is distributed according to
$t_{p,\nu}(\mu,\Psi)$; see \cite{kotz}, Chapter~1.  This scale-mixture
representation, illustrated in Figure \ref{fig:standard}, allows for
easy sampling.  It also clarifies how the use of $t$-distributions
leads to more robust inference because extreme observations can arise
from small values of $\tau$.  An additional useful fact is that the
conditional distribution of $\tau$ given $Y$ is again a
Gamma-distribution, namely,
\begin{equation}  \label{eq:tau-given-Y}
  (\tau\vert Y) \sim \Gamma\biggl( \frac{\nu + p}{2}, \frac{\nu +
    \delta_{Y}(\mu,\Psi)}{2}\biggr).
\end{equation}

Let $G=(V,E)$ be a graph with vertex set $V=\{1,\dots,p\}$. We define
the associated graphical model for the $t$-distribution by requiring
that $\theta_{jk}=0$ for indices $j\not=k$ corresponding to a nonedge
$(j,k)\notin E$.  This mimics the Gaussian model in that zero
constraints are imposed on the inverse of the covariance matrix.
However, in a $t$-distribution this no longer corresponds to
conditional independence, and the density $f_\nu(y;\mu,\Psi)$ does not
factor according to the graph. The conditional dependence manifests
itself, in particular, in conditional variances in that even if
$\theta_{jk}=0$,\vspace*{-1pt}
\[
\mathbb{V}[Y_j|Y_{\setminus j}]\not=
\mathbb{V}\bigl[Y_j|Y_{\setminus\{j,k\}}\bigr].\vspace*{-1pt}
\]
For a simple illustration of this inequality, let $\Psi$ be a diagonal
matrix.  Then\vspace*{-2pt}
\[
  \mathbb{V}[Y_j|Y_{\setminus j}] = \mathbb{E}[X_j^2/\tau|Y_{\setminus j}]
  = \frac{1}{\theta_{jj}}\cdot\mathbb{E}[\tau^{-1}|Y_{\setminus j}] =
  \frac{1}{\theta_{jj}}\cdot\frac{\nu +
    \delta_{Y_{\setminus j}}(\mu_{\setminus j},\Psi_{\setminus j,\setminus
      j})}{\nu + p-3},\vspace*{-2pt}
\]
which can be shown by taking iterated conditional expectations, and
using that\vspace*{-5pt}
\[
\mathbb{E}[X_j^2|Y_{\setminus j},\tau]= \mathbb{E}[X_j^2|X_{\setminus
j},\tau]= \mathbb{V}[X_j|X_{\setminus j}]=\frac{1}{\theta_{jj}}\vspace*{-2pt}
\]
and that $\tau$ given $Y_{\setminus j}$ has a Gamma-distribution;
recall (\ref{eq:tau-given-Y}).  Clearly, $ \mathbb{V}[Y_j|Y_{\setminus
  j}]$ depends on all $Y_k$, $k\not= j$.

Despite the lack of conditional independence, the following property
still holds (proved in the \hyperref[appendix]{Appendix}).\vspace*{-2pt}

\begin{proposition} \label{thm:1}
 Let $X \sim \mathcal{N}_p(0,\Theta^{-1})$, where $\theta_{jk}=0$ for
 pairs of indices $j\not= k$ that correspond to nonedges in the graph $G$.
 Let $\tau$ be independent of~$X$ and follow any distribution on the positive real numbers with $\mathbb{E}[1/\tau] < \infty$ and define
 $Y=\mu + X/\sqrt{\tau}$.  If two nodes $j$ and $k$ are separated by a set
 of nodes~$C$ in $G$, then $Y_j$ and $Y_k$ are conditionally uncorrelated
 given $Y_{C}$.\vspace*{-2pt}
 \end{proposition}

The edges in the graph indicate the allowed conditional independencies
in the latent Gaussian vector $X$.  According to Proposition
\ref{thm:1}, however, we may also interpret the graph in terms of the
observed variables $Y_j$.  The zero conditional correlations entail
that mean-square error optimal prediction of variable $Y_j$ can be
based on the variables $Y_k$ that correspond to neighbors of the node
$j$ in the graph, which is a very appealing property.\vspace*{-2pt}

\subsection{EM algorithm for estimation}\label{sec:em}

The lack of density factorization properties complicates likelihood
inference with $t$-distributions.  However, the EM algorithm provides a
way to circumvent this issue.  Equipped with the normal-Gamma
construction, we treat $\tau$ as a hidden variable and use that the
conditional distribution of $Y$ given $\tau$ is
$\mathcal{N}_p(\mu,\Psi/\tau)$.  We now outline the EM algorithm for
the $t$-distribution assuming the degrees of freedom $\nu$ to be known.
If desired, $\nu$ could also be estimated in a line search that is best
based on the actual $t$-likelihood [Liu and Rubin~\citeyear{liurubin}].

Consider an $n$-sample $Y_1,\dots,Y_n$ drawn from
$t_{p,\nu}(\mu,\Psi)$. Let $\tau_1,\dots,\tau_n$ be an associated
sequence of hidden Gamma-random variables.  Observation of the $\tau_i$
would lead to the following complete-data log-likelihood function for~%
$\mu$ and $\Theta=\Psi^{-1}$:\vspace*{-5pt}
\begin{eqnarray}
  \ell_{\mathrm hid}(\mu,\Theta |Y,\tau) &\propto& \frac{n}{2} \log\det(\Theta)
  - \frac{1}{2} \operatorname{tr}\Biggl(\Theta \sum_{i=1}^n \tau_i Y_i
  Y_i^T\Biggr)\nonumber\\ [-11pt]\\ [-11pt]
  &&{} + \mu^T \Theta \sum_{i=1}^n \tau_i Y_i - \frac{1}{2} \mu^T
  \Theta \mu \sum_{i=1}^n \tau_i,\nonumber\vspace*{-2pt}
\end{eqnarray}
where, with some abuse, the symbol $\propto$ indicates that irrelevant
additive constants are omitted.  The complete-data sufficient
statistics\vspace*{-2pt}
\[
S_{\tau} = \sum_{i=1}^n \tau_i,\qquad  S_{\tau Y} = \sum_{i=1}^n \tau_i
Y_i,\qquad  S_{\tau YY} = \sum_{i=1}^n  \tau_i Y_i Y_i^T\vspace*{-2pt}
\]
are thus linear in $\tau$.  We obtain the following EM algorithm for
computing the maximum likelihood estimates of $\mu$ and $\Psi$:
\begin{description}
\item[E-step:] The E-step is simple because\vspace*{-2pt}
\begin{equation}  \label{eq:etau}
  \mathbb{E}[\tau | Y] = \frac{\nu + p}{\nu + \delta_Y(\mu,\Psi)}.\vspace*{-2pt}
\end{equation}
Given current estimates $\mu^{(t)}$ and $\Psi^{(t)}$, we compute in the
$(t+1)$st iteration\vspace*{-2pt}
\[
\tau_i^{(t+1)} = \frac{\nu + p}{\nu+\delta_Y(\mu^{(t)},\Psi^{(t)})}.\vspace*{-2pt}
\]
\item[M-step:] Calculate the updated estimates\vspace*{-2pt}
\begin{eqnarray}\label{eq:mut+1}
  \mu^{(t+1)} &=& \frac{\sum_{i=1}^n \tau_i^{(t+1)}
  Y_i}{\sum_{i=1}^n \tau_i^{(t+1)}},\\[-1pt]
  \Psi^{(t+1)} &=& \frac{1}{n} \sum_{i=1}^n \tau_i^{(t+1)} \bigl[
  Y_i-\mu^{(t+1)}\bigr] \bigl[ Y_i-\mu^{(t+1)}\bigr]^T.\vspace*{-1pt}
\end{eqnarray}
\end{description}

\section{Penalized inference in $t$-distribution models}\label{sec:tlasso}

Model selection in graphical $t$-models can be performed, in principle,
by any of the classical constraint- and score-based methods.  In
score-based searches through the set of all undirected graphs on $p$
nodes, however, each model would have to be refit using an iterative
method such as the algorithm from Section \ref{sec:em}.  The penalized
likelihood approach avoids this problem.

Like in the Gaussian case, we put a one-norm penalty on the elements
of~$\Theta$ and wish to maximize the penalized log-likelihood function
\begin{equation}
  \ell_{\rho,\mathrm{obs}}(\mu,\Theta|Y) =  \sum_{i=1}^n \log
  f_\nu(Y_i;\mu,\Theta^{-1}) -
  \rho\|\Theta\|_1,
\end{equation}
where $f_\nu$ is the $t$-density from (\ref{eq:tdensity}).  To achieve
this, we will use a modified version of the EM algorithm taking into
account the one-norm penalty.

We treat $\tau$ as missing data.  In the E-step of our algorithm, we
calculate the conditional expectation of the penalized complete-data
log-likelihood
\begin{equation}\label{pcdl}
  \ell_{\rho,\mathrm{hid}}(\mu,\Theta | Y,\tau)
  \propto \frac{n}{2} \log |\Theta| - \frac{n}{2} \operatorname{tr} (\Theta
  S_{\tau  YY}(\mu)) -\rho \|\Theta\|_1
\end{equation}
with
\[
  S_{\tau YY}(\mu) = \frac{1}{n} \sum_{i=1}^n \tau_i (Y_i -
  \mu)(Y_i- \mu)^T.
\]
Since $\ell_{\rho,\mathrm{hid}}(\mu,\Theta | Y,\tau)$ is again linear in
$\tau$, the E-step takes the same form as in Section \ref{sec:em}.  Let
$\mu^{(t)}$ and $\Theta^{(t)}$ be the estimates after the $t$th
iteration, and~$\tau_i^{(t+1)}$ the conditional expectation of $\tau_i$
calculated in the $(t+1)$st E-step.  Then in the M-step of our
algorithm we wish to maximize
\[
  \frac{n}{2} \log |\Theta| - \frac{n}{2} \operatorname{tr} \bigl(\Theta
  S_{\tau^{(t+1)} YY}(\mu)\bigr)  -\rho \|\Theta\|_1
\]
with respect to $\mu$ and $\Theta$.  Differentiation with respect to
$\mu$ yields $\mu^{(t+1)}$ from~(\ref{eq:mut+1}) for any value of
$\Theta$. Therefore, $\Theta^{(t+1)}$ is found by maximizing
\begin{equation}  \label{eq:hidglasso}
  \frac{n}{2} \log |\Theta| - \frac{n}{2} \operatorname{tr} \bigl(\Theta S_{\tau^{(t+1)}
    YY}\bigl(\mu^{(t+1)}\bigr)\bigr) -\rho \|\Theta\|_1.
\end{equation}
The quantity in (\ref{eq:hidglasso}), however, is exactly the objective
function maximized by the \textit{glasso}.

Iterating the E- and M-steps just described, we obtain what we call the
\textit{tlasso} algorithm.  Since the one-norm penalty forces some elements of
$\Theta$ exactly to zero, the \textit{tlasso} performs model selection
and parameter estimation in a way that is similar to structural EM
algorithms [\cite{friedman}].  Convergence to a stationary point is
guaranteed in the penalized version of the EM algorithm
[\cite{mclachlan}, Chapter 1.6]; typically a~local maximum is found.
Note also that the maximized log-likelihood function is not concave,
and so one finds oneself in the usual situation of not being able to
give any guarantees about having obtained a global maximum.

\section{Alternative model}\label{sec:alasso}%

\subsection{Specification of the alternative $t$-model}

The \textit{tlasso} from Section \ref{sec:tlasso} performs particularly
well when a small fraction of the observations are contaminated (or
otherwise extreme).  In this case, these observations are downweighted
in entirety, and the gain from reducing the effect of contaminated
nodes outweighs the loss from throwing away good data from other nodes.
In high-dimensional data sets, however, the contamination, or other
deviation form normality, may be in small parts of many observations.
Downweighting entire observations may then no longer achieve the
desired results.  We will demonstrate this later in simulations (see
the bottom panel of Figure \ref{fig:sims}).

\begin{figure}

\includegraphics{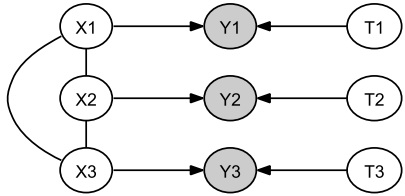}

\caption{Graph representing the process generating a $t^*$-random
vector $Y$ from a latent Gaussian random vector $X$ and independent latent
  Gamma-divisors.}\label{fig:alternative}
\end{figure}

To handle the above situation better, we consider an alternative
extension of the univariate $t$-distribution, illustrated in
Figure \ref{fig:alternative}.  Instead of one divisor~$\tau$ per
$p$-variate observation, we draw $p$ divisors $\tau_j$.  For $j=1,
\ldots ,p$, let $\tau_j \sim \Gamma(\nu/2,\nu/2)$ be independent of
each other and of $X \sim \mathcal{N}_p(0,\Psi)$.  We then say that the
random vector $Y$ with coordinates $Y_j = \mu_j + X_j/\sqrt{\tau_j}$
follows an alternative multivariate $t$-distribution; in symbols $Y
\sim t^*_{p,\nu}(\mu,\Psi)$.

Unlike for the classical multivariate $t$-distribution, the covariance
matrix~$\mathbb{V}[Y]$ is no longer a constant multiple of
$\Psi=(\psi_{jk})$ when $Y \sim t^*_{p,\nu}(\mu,\Psi)$.  Clearly, the
coordinate variances are still the same, namely,
\[
\mathbb{V}[Y_j]=\frac{\nu}{\nu-2} \cdot \psi_{jj},
\]
but the covariance between $Y_j$ and $Y_k$ with $j\not=k$ is now
\[
  \frac{\nu\Gamma((\nu -1)/2)^2}{2\Gamma(\nu/2)^2}
  \cdot\psi_{jk} \le
\frac{\nu}{\nu-2} \cdot \psi_{jk}.
\]
The same matrix $\Psi$ thus implies smaller correlations (by the same
constant multiple) in the $t^*$-distribution.  This reduced dependence
is not surprising in light of the fact that now different and
independent divisors appear in the different coordinates.  Despite the
decrease in marginal correlations, the result of
Proposition \ref{thm:1} does not hold for conditional correlations in
the alternative model.  That is, $\Psi^{-1}_{jk}=0$ does not imply
$Y_j$ and $Y_k$ are conditionally uncorrelated given
$Y_{\setminus\{j,k\}}$.  Interpretation of the graph in the alternative
model is thus limited to considering edges to represent the allowed
conditional dependencies in the latent multivariate normal
distribution.

\begin{figure}

\includegraphics{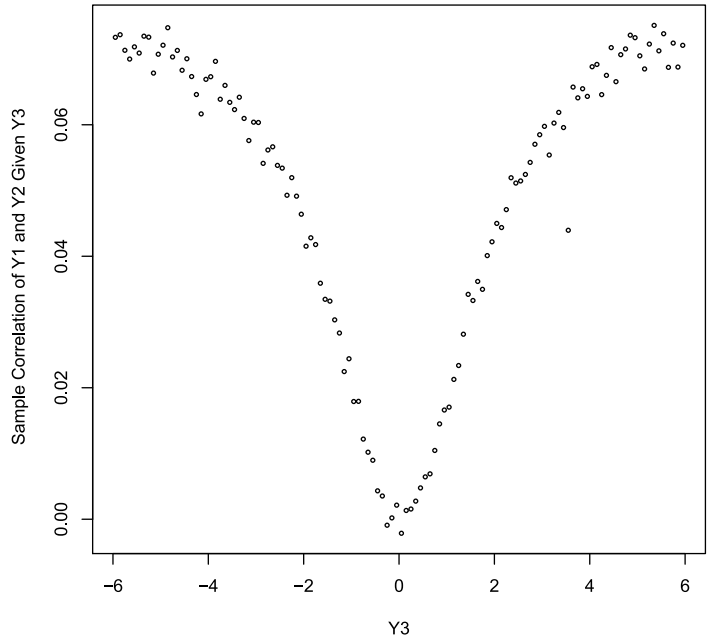}
\vspace*{-5pt}
\caption{Sample correlation of $Y_1$ and $Y_2$ for observations with
$Y_3$ in a window of size 0.01.}\label{fig:condcor}\vspace*{-5pt}
\end{figure}

The following simulation confirms the result and illustrates the
effect. We consider a $t^*_{3,3}(0,\Theta^{-1})$ distribution with
\[
\Theta =
\left[\matrix{
1&0&-0.5\cr
0&1&-0.5\cr
-0.5&-0.5&1\cr
}\right]
\]
and draw independent samples until we have 500,000 observations with
$x < Y_3 < x +0.01$ for $120$ values of $x$ in the range $(-6,6)$.  The
sample correlations of $Y_1$ and $Y_2$ given the varying values of
$Y_3$ are shown in Figure \ref{fig:condcor}.

\subsection{Alternative tlasso}

Inference in the alternative model presents some difficulties because
the likelihood function is not available explicitly.  The complete-data
log-likelihood function $\ell^*_{\rho,\mathrm{hid}}(\mu,\Theta|Y,\tau)$,
however, is simply the product of the evaluations of $p$
Gamma-densities ($\tau$ being a vector now) and a multivariate normal
density.  We can thus implement an EM-type procedure if we are able to
compute the conditional expectation of $\ell^*_{\rho,\mathrm{hid}}(\mu,\Theta|Y,\tau)$ given $Y=(Y_1,\dots,Y_n)$.  This time we treat
the $p$ random variables $(\tau_{i1}, \ldots ,\tau_{ip})$ as hidden for
each observation $i=1,\dots,n$.  Unfortunately, the conditional
expectation is intractable. It can be estimated, however, using Markov
Chain Monte Carlo.

The complete-data log-likelihood function is equal to
\begin{equation}\label{altpcdl}
  \ell^*_{\rho,\mathrm{hid}} (\mu , \Theta, | Y, \tau)
  \propto \frac{n}{2} \log |\Theta|- \frac{n}{2} \operatorname{tr} (\Theta
  S^*_{\tau YY}(\mu) )
  -\rho \|\Theta\|_1,
\end{equation}
where
\[
  S^*_{\tau YY}(\mu) = \frac{1}{n} \sum_{i=1}^n
  D\bigl(\sqrt{\tau_i}\bigr) (Y_i - \mu)(Y_i- \mu)^TD\bigl(\sqrt{\tau_i}\bigr)
\]
and $D(\sqrt{\tau_i})$ is the diagonal matrix with
$\sqrt{\tau_i}=\sqrt{\tau_{i1}}, \ldots ,\sqrt{\tau_{ip}}$ along the
diagonal.  The trace in (\ref{altpcdl}) is linear in the entries of the
matrix $\sqrt{\tau_i}\sqrt{\tau_i}^T$.  A Gibbs sampler for estimating
the conditional expectation of this matrix given $Y$ cycles through the
coordinates indexed by $j=1, \ldots ,p$ and draws, in its $m$th
iteration, a number $\tau_{ij}^{(m)}$ from the conditional distribution
of $\tau_{ij}$ given $(\tau_{i \setminus j},Y)$. This full conditional
has density
\begin{equation} \label{eq:full-conditional}
  f(\tau_{ij}|\tau_{i \setminus j},Y_i) = C(\alpha,\beta,\gamma)\cdot
  \tau_{ij}^{\alpha
  - 1} \exp \bigl\{-\tau_{ij} \beta - \sqrt{\tau_{ij}} \gamma \bigr\}
\end{equation}
with
\begin{equation}\label{eq:exactgibbs}
 \hspace*{30pt}\alpha = \frac{\nu+1}{2}, \qquad \beta = \frac{\nu +
(Y_{ij}-\mu_j)^2\theta_{jj}}{2}, \qquad \gamma = (Y_{ij}-\mu_j)\Theta_{j
\setminus j}X_{ i\setminus j},
\end{equation}
and normalizing constant $C(\alpha,\beta,\gamma)$.  This constant can
always be expressed using hypergeometric functions, but, as we detail
below, much simpler formulas can be obtained for the small integer
degrees of freedom $\nu$ that are of interest in practice.  The simpler
formulas are obtained by partial integration.

From $\beta$ and $\gamma$ in (\ref{eq:exactgibbs}), form the ratio
$\gamma'=\gamma/(2\sqrt{\beta})$.  In order to sample from the
distribution in (\ref{eq:full-conditional}), we may draw from
\begin{equation} \label{eq:full-cond-beta11}
  f_{\alpha,\gamma'}(t) = C(\alpha,\gamma')\cdot
  t^{\alpha
  - 1} \exp \bigl\{-t - \sqrt{t} 2\gamma' \bigr\}
\end{equation}
and divide the result by $\beta$.  For our default of $\nu=3$, that is,
$\alpha=2$, we thus need to sample from
\begin{equation} \label{eq:full-cond-beta1}
  f_{\gamma}(t) = C(\gamma)\cdot
  t \exp \bigl\{-t - \sqrt{t} 2\gamma \bigr\} .
\end{equation}
Writing $\Phi$ for the cumulative distribution function of the standard
normal distribution, the normalizing constant becomes
\begin{equation}\label{eq:erfc}
  1/C(\gamma) = 1+
  \gamma^2-\gamma(2\gamma^2+3)\sqrt{\pi}\exp\{\gamma^2\}\bigl(1- \Phi
    \bigl(\gamma\sqrt{2}\bigr)\bigr).
\end{equation}
For $\gamma=0$, the density $f_\gamma(t)$ is a $\Gamma(2,1)$ density.
For moderate $\gamma$, we are thus led to the following rejection
sampling procedure to draw from $f_\gamma$.

Let $g_{\delta}$ be the density of a $\Gamma(2,\delta)$ distribution.
Rejection sampling using the family of densities $g_{\delta}$ as
instrumental densities proceeds by drawing a~proposal $T \sim
\Gamma(2,\delta)$ and a uniform random variable $U \sim
\mathcal{U}(0,1)$ and either accept if $U \le
f(T)/(M_{\delta}g_{\delta}(T))$ or repeat the process until acceptance.
Here,~$M_{\delta}$ is a suitable multiplier such that $f(t) \le
M_{\delta} g_{\delta}(t)$ for all $t\ge 0$.

An important ingredient to the rejection sampler is the parameter
$\delta$, which we choose as follows.  In the case $\gamma < 0$, the
density $g_{\delta}$ has a heavier tail than~$f$ provided that
$\delta<1$. Focusing on the case $\alpha=2$, we have that for a given
$\delta<1$ the smallest $M$ such that $f(t) \le M g_{\delta}(t)$ for
all $t\ge 0$ is
\[
M_{\delta}= C \cdot\frac{1}{\delta^{2}} \exp \biggl\{
\frac{\gamma^2}{(1-\delta)} \biggr\}.
\]
Varying $\delta$, the multiplier $M_{\delta}$ is minimized at
\[
\delta= 1 + \frac{\gamma^2 - \sqrt{\gamma^4+8
  \gamma^2}}{4}.
\]
If $\gamma>0$, then\vadjust{\goodbreak} setting $\delta=1$ yields a heavy enough tail and
$M_{\delta}=C$.

The rejection sampling performs draws from the exact conditional
distribution $f(\tau_{ij}|\tau_{i \setminus j},Y)$.  We find it works
very well for data with not too extreme contamination such as, for
instance, in the original as well as bootstrap data from the
application discussed in Section \ref{sec:isoprenoid-pathway}.  When
applied to data with very extreme observations $Y_{ij}$, however, one
is faced with larger positive values of $\gamma$.  In this case the
instrumental densities $g_\delta$ provide a poor approximation to the
target density $f_\gamma$, and the acceptance probabilities in the
rejection sampling step become impractically low.

For $\gamma >1$, we thus use an alternative rejection procedure.  Make
the transformation $s=\sqrt{t}$.  We then wish to sample from
\[
h_{\gamma}(s) = 2 C(\gamma)\cdot s^3 \exp  \{-s^2 - s 2\gamma
\} .
\]
Any $\Gamma(\alpha,\delta)$ distribution has a heavier tail than the
target distribution $h_{\gamma}(s)$.  While it is not possible to find
an analytical solution for the optimal $\alpha$ and~$\delta$, letting
$\alpha=1$ and $\delta=(\gamma+1)/2$ yields acceptance probabilities
between $40\%$ and $50\%$ for most plausible values of $\gamma$.  Since
this alternative procedure will only be needed occasionally, these
acceptance problems are adequate.  Using this hybrid approach yields
overall acceptance probabilities greater than $98\%$ for the data with
extreme contamination described in Section \ref{sec:galactose}.

Returning to the iterations of the overall sampler, we calculate
$\sqrt{\tau_i}\sqrt{\tau_i}^T$ at the end of each cycle through the $p$
nodes, and then take the average over~$M$ iterations.  This solves the
problem of carrying out one E-step, and we obtain the following
stochastic penalized EM algorithm, which we call the Monte Carlo
$t^*$\textit{-lasso} (or $t^*_{\mathrm MC}$\textit{-lasso} for short):
\begin{description}
\item[E-step:] Given current estimates $\mu^{(t)}$ and $\Psi^{(t)}$,
compute ${(\sqrt{\tau_i}\sqrt{\tau_i}^T)}^{(t+1)}$ by averaging the matrices
  obtained in some large number $M$ of Gibbs sampler iterations, as
  described above.
\item[M-step:] Calculate the updated estimates
\[
  \mu_j^{(t+1)} = \frac{\sum_{i=1}^n \tau_{ij}^{(t+1)}
    Y_{ij}}{\sum_{i=1}^n \tau_{ij}^{(t+1)}}.
\]
Use these and ${(\sqrt{\tau_i}\sqrt{\tau_i}^T)}^{(t+1)}$ to compute the
matrix $S^*_{\tau^{(t+1)} YY}(\mu^{(t+1)})$ to be plugged into the
trace term in (\ref{altpcdl}).  Maximize the resulting penalized
log-likelihood function using the \textit{glasso}.
\end{description}

\subsection{Variational approximation}\label{sec:variational}
 The above Monte Carlo procedure loses much of
the computational efficiency of the classical \textit{tlasso} from
Section~\ref{sec:tlasso}, however, and can be prohibitively expensive
for large $p$.  For large problems, we turn instead to variational
approximations of the conditional densi\-ty~$f (\tau_i|Y_i)$ of the
vector $\tau_i$ given the observed vector $Y_i$.

The variational approach proceeds by approximating the conditional
density $f(\tau_{ij}|Y_i)$ by a factoring distribution.  In our
context, however, it is easier to approximate the joint density
$f(\tau_i,Y_i)=f(\tau_i)f(Y_i|\tau_i)$ instead.  The first term is
already in product form because we are assuming the individual divisor
$\tau_{ij}$ to be independent in the model formulation, and the second
term is the density of the multivariate normal distribution
\[
\mathcal{N}_p\bigl(\mu,D\bigl(1/\sqrt{\tau_i}\bigr)\Theta^{-1}
  D\bigl(1/\sqrt{\tau_i}\bigr)\bigr).
\]
We approximate this normal distribution by a member of the set of
multivariate normal distributions with diagonal covariance matrix.
Application of this naive mean field procedure, that is, choosing a
distribution by minimizing Kullback--Leibler divergence, leads to the
approximating distribution
\begin{equation}  \label{eq:meanfield}
  \mathcal{N}_p\bigl(\mu,D\bigl(1/\sqrt{\tau}\bigr)\bar{\Theta}^{-1} D\bigl(1/\sqrt{\tau}\bigr)\bigr),
\end{equation}
where $\bar{\Theta}$ is the diagonal matrix with the same diagonal
elements as $\Theta$ [\cite{wainwrightjordan}, Chapter 5].  Writing
$q^*(Y|\tau)$ for the density of the distribution in
(\ref{eq:meanfield}), our approximation thus has the fully factoring
form $q^*_{\tau_i,Y_i}(\tau_i,Y_i)=f(\tau_i)q^*(Y_i|\tau_i)$.  The
resulting conditional distribution also factors as
\[
q^*(\tau_i|Y_i)= \prod_{j=1}^p g(\tau_{ij}|Y_{ij}),
\]
where $g(\tau_{ij}|Y_{ij})$ is the density of the Gamma-distribution
$\Gamma(\alpha_{ij},\beta_{ij})$, with its parameters corresponding to
the quantities $\alpha$ and $\beta$ in (\ref{eq:exactgibbs}).

In conclusion, the variational E-step consists of calculating, for each
observation $Y_i$, the expectations
\[
  \mathbb{E}_g[\tau_{ij} | Y_{ij}] =
  \frac{\alpha_{ij}}{\beta_{ij}}, \qquad
  \mathbb{E}_g\bigl[\sqrt{\tau_{ij}} | Y_{ij}\bigr] =
  \frac{\Gamma(\alpha_{ij}+1/2)}{\Gamma(\alpha_{ij})\sqrt{\beta_{ij}}},
\]
and $\mathbb{E}[\sqrt{\tau_j}\sqrt{\tau_k} | Y_{i} ] =
\mathbb{E}_g[\sqrt{\tau_j}|Y_{ij}] \mathbb{E}_g[\sqrt{\tau_k} | Y_{ik}
]$. These values\vspace*{1pt} are then substituted into (\ref{altpcdl}).  The M-step
is the same as in the $t^*_{\mathrm{MC}}$\textit{-lasso}.

The effect of the variational approximation is that the weight for
node~$j$ in observation $i$ is based solely on the squared deviation from
the mean, $(Y_{ij} - \mu_j)^2$ and the conditional variance
$1/\theta_{jj}$.  For a given deviation from the mean, the larger the
conditional variance of the node, the smaller the weight given to that
node in that observation.  But unlike in the $t^*_{\mathrm{MC}}$\textit{-lasso},
 no consideration is given to deviation from the conditional
mean of the node in question given the rest. Some relevant information
is therefore not being used, but in our simulations the effect was not
noticeable.

The resulting variational $t^*$\textit{-lasso} ($t^*_{\mathrm{var}}$\textit{-lasso})
 is only slightly more expensive than the \textit{tlasso} and,
despite the relatively crude approximation in the variational E-step,
performs well compared with the $t^*_{\mathrm{MC}}$\textit{-lasso}.  Because
of this, we will use exclusively the $t^*_{\mathrm{var}}$\textit{-lasso}
when considering the alternative model in the simulations in the next
section.

\section{Simulation results}\label{sec:simulations}
\subsection{Procedure}
We used simulated data to compare the three procedures \textit{glasso},
\textit{tlasso} and $t^*_{\mathrm{var}}$\textit{-lasso} as follows.  We generated a random $100\times100$
sparse inverse covariance (or dispersion) matrix $\Theta$ according to
the following procedure:
\begin{enumerate}[(a)]
\item[(a)] Choose each lower-triangular element of $\Theta$
independently to be
  $-1$, $0$ or $1$ with probability $1\%$, $98\%$ and $1\%$, respectively.
\item[(b)] For $j>k$ set $\theta_{kj}=\theta_{jk}$. \item[(c)] Define
$\theta_{kk}=1+h$ where $h$ is the number of nonzero elements in
  the $k$th row of $\Theta$.
\end{enumerate}
The final step ensures a strictly diagonally dominant, and thus
positive-definite matrix.  To strengthen the partial correlations, we
reduced the diagonal elements by a common factor.  We made this factor
as large as possible while maintaining positive-definiteness and
stability for inversion.  For these particular matrices, fixing a
minimum eigenvalue of $0.6$ worked well.

We then generated $n=50$ observations from the
$\mathcal{N}_{100}(0,\Theta^{-1})$ distribution and ran each of the
three procedures with a range of values for the one-norm tuning
parameter $\rho$. To compare how well the competing methods recovered
the true edges, we drew ROC curves.  We ran this whole process 250
times and then repeated the entire computation, drawing data from
$t_{100,3}(0,\Theta^{-1})$ and then $t^*_{100,3}(0,\Theta^{-1})$
distributions.

Simulating from $t$-distributions produces extreme observations, but a
more realistic setting might be one in which normal data is
contaminated in some fashion.  For instance, consider broken probes or
misread observations in a~large gene expression microarray.  Suppose
the contaminated data are not so extreme as to be manually screened or
otherwise identified as obvious outliers. To simulate this phenomenon,
we generate normal data as above, but randomly contaminated $2\%$ of
the values with data generated from independent univariate
$\mathcal{N}(\mu^*,0.2)$ random variables, where $\mu^*$ is equal to
$2.5$ times the largest diagonal element of $\Theta^{-1}$.  These
contaminated values will be similar in magnitude to the $2\%$ tail of
the original $\mathcal{N}_{100}(0,\Theta^{-1})$ distribution and
therefore difficult to identify.

Finally, we would like to compare our developed $t$-procedures with
simpler approaches to robust inference.  There are many ways to obtain
robust estimates of the covariance matrix, but these usually require
$n>p$.  Instead we obtain a robust estimate for the marginal
covariances and variances using the procedure of \cite{kalisch}.  Since
this is not guaranteed to result in a positive definite matrix, we add
a constant, $c$, to the diagonal elements of the matrix, where $c$ is
the minimum constant necessary to ensure the resulting matrix is
nonnegative definite.  We then use this robust estimate of the
covariance matrix as input into the \textit{glasso} and refer to this
procedure as the \textit{robust glasso}.

 \subsection{Results}

Our \textit{tlasso} and $t^*_{\mathrm{var}}$\textit{-lasso} are
computationally more expensive, since they call the \textit{glasso} at
each M-step.  But in our simulations, the algorithms converge quickly.
If we run through multiple increasing values of the tuning parameter
$\rho$ for the one-norm penalty, it may take about $15$--$30$ EM
iterations for the initial small value of $\rho$, but only 2 or 3
iterations for later values, as we can ``warm start'' at the previous
output.  But even in the initial run, two iterations typically lead to
a drastic improvement (in the $t$ likelihood) over the \textit{glasso}.

The only caveat is that the function being maximized by the \textit{tlasso}
 methods is not guaranteed to be unimodal.  We thus started in
several places, and let the algorithm run for longer than probably
necessary in practice.  We did not observe drastically different
results from different starting places.  Nonetheless, since we are not
guaranteed to find a global maximum, the statistical performances of
the \textit{tlasso} and $t^*_{\mathrm{var}}$\textit{-lasso} may, in principle, be understated here (and, of course,
the computational efficiency overstated).

\begin{figure}

\includegraphics{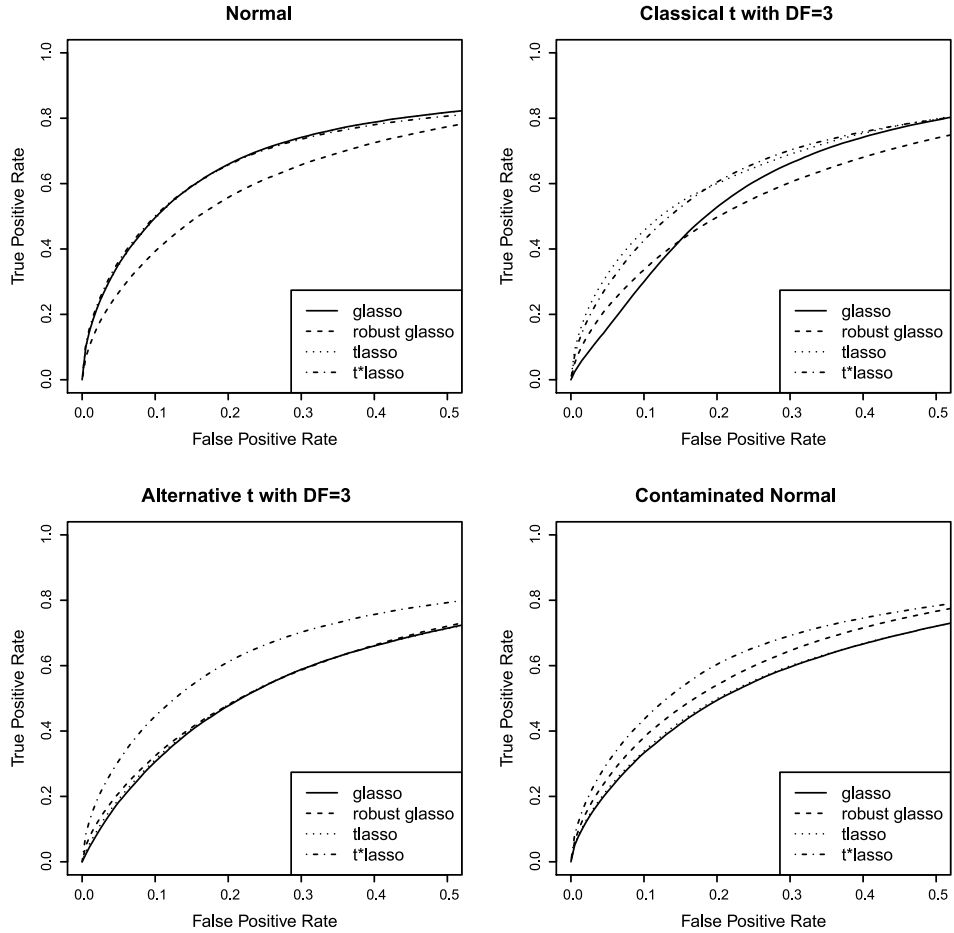}
\vspace*{-5pt}
\caption{ROC curves depicting the performances of the four methods
under four different types of data.  Each curve is an average over 250
simulations.} \label{fig:sims}\vspace*{-6pt}
\end{figure}

In the worst case scenario for our procedures relative to the \textit{glasso}---when
 the data is normal and we assume $t$-distributions with
$3$ degrees of freedom---almost no statistical efficiency is lost.  In
the numerous simulations we have run using normal data, the \textit{tlasso} and
 \textit{glasso} do an essentially equally good job of
recovering the true graph (see Figure \ref{fig:sims}).  The $t^*_{\mathrm{var}}$\textit{-lasso} performs
surprisingly well at small to moderate false discovery rates.  The
 \textit{robust glasso} is based on a less efficient estimator and does not
perform as well as the other procedures.

For data generated from a classical $t$-distribution with $3$ degrees
of freedom, the \textit{tlasso} provides drastic improvement over the
\textit{glasso} at the low false positive rates that are of practical interest.
The assumed normality and the occasional extreme observation lead to
numerous false positives when using the \textit{glasso}.  Therefore,
there is very little computational---and little or no
statistical---downside to assuming $t$-distributions, but significant
statistical upside. Interestingly, the $t^*_{\mathrm{var}}$\textit{-lasso}
performs about as well as the \textit{tlasso}.  The \textit{robust
glasso} outperforms the purely Gaussian procedure at low false positive
rates, since it is less susceptible to the most extreme
observations.

In the third case, with data generated from the alternative
$t$-distribution with $3$~degrees of freedom, only the $t^*_{\mathrm{var}}$\textit{-lasso}
 is able to recover useful information without substantial noise.
The occasional large values are too extreme for the normal model to
explain and downweighting entire observations, as is done by the \textit{tlasso},
 discards too much information when there are extreme values
scattered throughout the data.  The \textit{robust glasso} offers only
a small improvement over the \textit{glasso}.

With the contaminated data, the $t^*_{\mathrm{var}}$\textit{-lasso} does not
perform as well in this case as it does with $t^*$ data.  The extreme
values are not downweighted as much and, thus, the signals are noisier.
It still performs far better, however, than either of the other
methods, and is able to recover valuable information in a case where
manual\ screening of outliers would be very difficult.  The
\textit{robust glasso} does not perform as well as the $t^*_{\mathrm{var}}$\textit{-lasso},
 but offers a clear improvement over the \textit{glasso}
 and might be a useful alternative.

\subsection{Notes on simulation}

The simulations show that the \textit{tlasso} performs very similarly
to the \textit{glasso} even with normal data.  While one would expect a~model based on the $t$-distribution to fare better with normal data
than a~normal model would with $t$ data, the fact that there is almost
no statistical loss from the model misspecification is at first a bit
surprising.  The similarity of the results can be explained, however,
by comparing the two procedures. In effect, the only difference is that
the \textit{tlasso} inputs a weighted sample covariance matrix into the
\textit{glasso} procedure; one can then think of the \textit{glasso} as
the \textit{tlasso}\vadjust{\goodbreak} with all weights set to~one.

As noted in Section \ref{sec:em}, these weights are the conditional
expectations of~$\tau$, which are, from equation (\ref{eq:etau}),
\begin{equation}
 \tau_i^{(t+1)}= \mathbb{E}[\tau_i | Y_i] = \frac{\hat{\nu} + p}{\hat{\nu}
 + \delta_{Y_i}(\mu^{(t)},\Psi^{(t)})},
\end{equation}
where $\hat{\nu}$ is our estimate or assumption of the unknown degrees
of freedom.  If $Y \sim t_{p,\nu}(\mu,\Psi)$ and $\nu > 4$, then
$\delta_Y(\mu,\Psi)/p$ is distributed according to the~%
$\mathcal{F}_{p,\nu}$ distribution [\cite{kotz}, Chapter 3].  Thus,
starting with the true values of $\mu$ and $\psi$, the variance of the
inverse weights is
\[
\mathbb{V} \biggl[ \frac{\hat{\nu} + \delta_Y(\mu,\Psi)}{\hat{\nu} + p}
\biggr]= \frac{2 p \nu^2 (p+\nu -2)}{(\nu-2)^2(\nu-4)(\hat{\nu}+p)^2}.
\]
For normal data (i.e., $\nu=\infty$), the variance is
$2p/(\hat{\nu}+p)^2$ and goes to $0$ very quickly as $p$ gets large, no
matter the assumed value of $\hat{\nu}$.  If our current estimate of
$\Theta$ is reasonably close to the true $\Theta$, then the
observations will likely have very similar weights and the weighted
covariance matrix will be very close to the sample covariance matrix.
For $t$ data, the above variance tends to $2 \nu^2/(\nu-2)^2(\nu-4)$
for large $p$; so no matter how many variables we have, the
distribution of the inverse weights will have positive variance and the
\textit{tlasso} and \textit{glasso} estimates are less likely to agree.

\section{Gene expression data} \label{sec:gene-expression-data}
\subsection{Galactose utilization}\label{sec:galactose}

We consider data from microarray experiments with yeast strands
[\cite{gasch}].  As in \cite{drtonrichardson}, we limit this
illustration to $8$ genes involved in galactose utilization.  An
assumption of normality is brought into question, in particular, by the
fact that in $11$ out of $136$ experiments with data for all 8 genes,
the measurements for 4 of the genes were abnormally large negative
values.  In order to assess the impact of this handful of outliers, we
run each algorithm, adjusting the penalty term $\rho$ such that a graph
with a given number of edges is inferred. Somewhat arbitrarily we focus
on the top 9 edges.  We do this once with all 136 experiments and then
again excluding the $11$ potential outliers.

\begin{figure}

\includegraphics{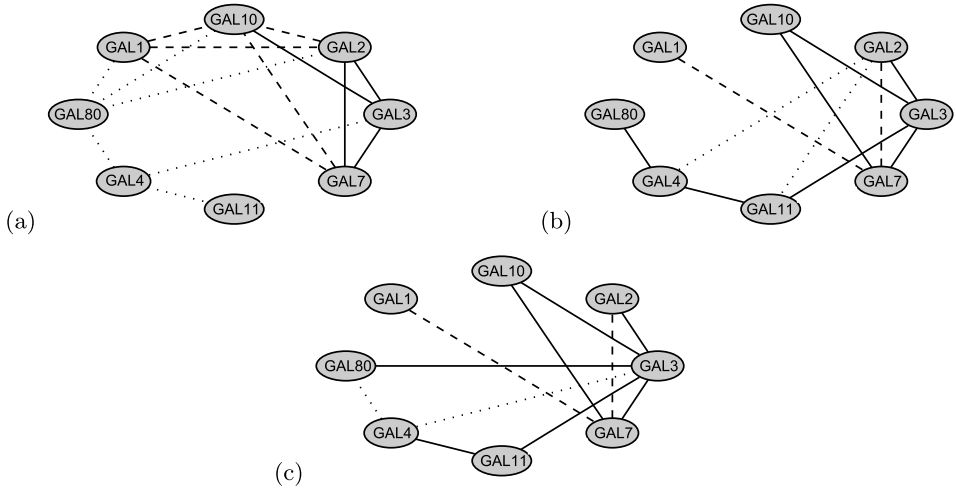}

\caption{Top 9 recovered edges: \textup{(a)} \textit{glasso},
\textup{(b)} \textit{tlasso}, \textup{(c)} $t^*_{\mathrm{MC}}$\textit{-lasso}.  Dashed
  edges were recovered only when including the outliers; dotted only when
  excluding them; solid in both cases.}\label{fig:gascht9}
\end{figure}

As seen in Figure \ref{fig:gascht9}, the \textit{glasso} infers very
different graphs, with only 3 edges in common.  When the ``outliers''
are included, the \textit{glasso} estimate in
Figure \ref{fig:gascht9}(a) has the 4 nodes in question fully
connected; when they are excluded, no edges among the 4 nodes are
inferred.  The \textit{tlasso} does not exhibit this extreme behavior.
As seen in Figure \ref{fig:gascht9}(b), it recovers almost the same
graph in each case (7 out of 9 edges shared).  When run with all the
data, the $\tau$ estimate is very small (${\sim}0.04$) for each of the
$11$ questionable observations compared with the average $\tau$
estimate of~$1.2$.\vadjust{\goodbreak}  The graph in Figure \ref{fig:gascht9}(c) shows the
results from the $t^*_{\mathrm{MC}}$\textit{-lasso} which performs just as
well as the \textit{tlasso}. The $t^*_{\mathrm{var}}$\textit{-lasso} also
recovered 7 edges in both graphs (not shown) and infers relationships
similar to those found by the $t^*_{\mathrm{MC}}$\textit{-lasso}.

\begin{figure}

\includegraphics{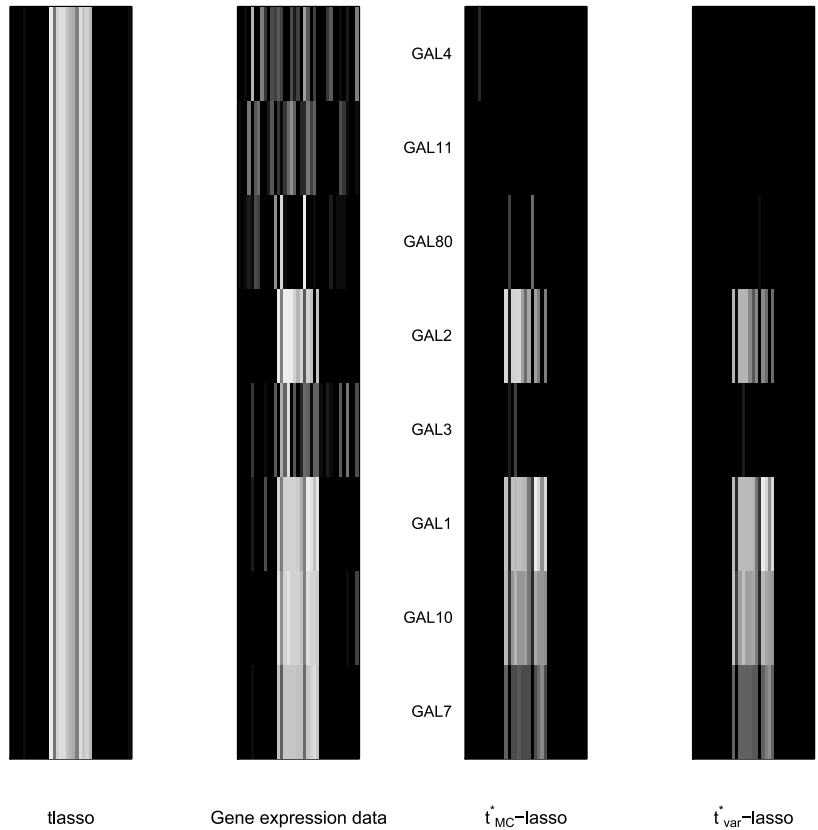}
\vspace*{-5pt}
\caption{From left to right, inverse weights from \textit{tlasso},
followed by normalized gene expression data, and inverse weights from $t^*_{\mathrm{MC}}$\textit{-lasso} and $t^*_{\mathrm{var}}$\textit{-lasso}.  Rows correspond to
  genes and columns to observations.  Lighter shades indicate larger values.
  The \textit{tlasso} uses only one weight per observation and so must weight
  each gene the same. All plots show the same subset of data including $11$
  potential outliers.}\label{fig:heatmap}\vspace*{-5pt}
\end{figure}

Figure \ref{fig:heatmap} illustrates the flexibility of the weighting
schemes of the various procedures.  Both $t^*$ procedures downweight
the 11 potential outliers observations for the 4 nodes in question, but
not for the other nodes.  Thus, the alternative version is able to
extract information from the ``uncontaminated'' part of the $11$
observations while downweighting the rest.  In this particular case,
with 125 other observations, downweighting the outliers is of primary
importance, and, thus, the increased flexibility of the $t^*_{\mathrm{MC}}$\textit{-lasso} over the \textit{tlasso} does not make much of a difference in the inferred graphs.  This
might not be the case with a higher contamination level.

\subsection{Isoprenoid pathway}\label{sec:isoprenoid-pathway}

We next consider gene expression data for the isoprenoid pathways of
\textit{
  Arabidopsis thaliana} discussed in \cite{wille}.  Gene expressions were
measured in 118 Affymetrix microarrays for 39 genes.  While the data
set described in the above section had clear deviations from normality,
the data described in this section has no obvious deviations that stand
out in exploratory plots.

\begin{figure}[b]

\includegraphics{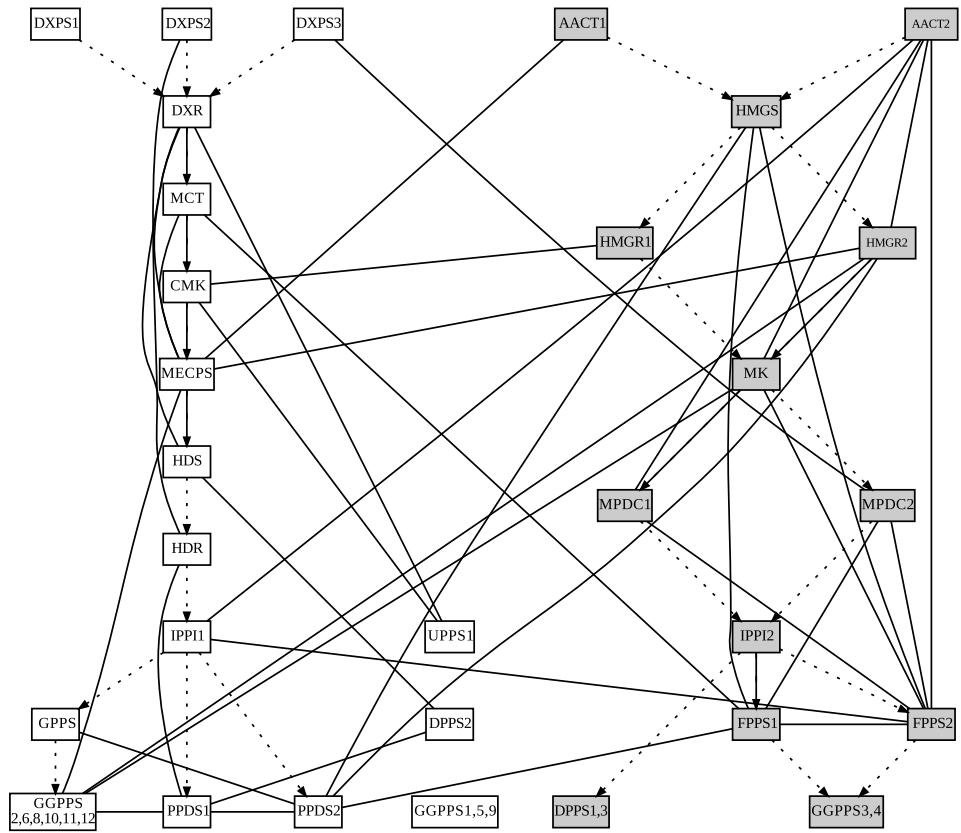}

\caption{A reproduction of the graph produced by Wille
et al. Solid
  undirected edges are those found by the model selection procedure; dotted
  arrows show the metabolic pathway.}\label{fig:isoprenoid1}\vspace*{3pt}
\end{figure}

Two approaches were considered in \cite{wille}.  The first
(\textit{GGM1}) fit a~Gaussian graphical model using BIC and backward
selection to obtain a network with 178 edges.  This number was deemed
too large for interpretation, and the authors considered instead only
the 31 edges found in at least $80\%$ of bootstrapped samples.  The
second approach (\textit{GGM2}) tests the conditional independence of each pair of genes given a
third gene.  An edge is drawn only if a test of conditional
independence is rejected for each other gene in the network.  This
approach is advocated in the paper and appears to find a network with
better biological interpretation.  The graph is shown in
Figure \ref{fig:isoprenoid1}, where shaded nodes indicate the so-called
MEP pathway.

Our approach is modeled after \textit{GGM1}.  We used the $t^*_{\mathrm{var}}$\textit{-lasso} and increasing values or $\rho$ to find a path of models to test.
For each chosen model, we ran the $t^*_{\mathrm{var}}$\textit{-lasso} again,
but this time without penalty on the allowed edges.  Since the $t^*$
likelihood is unavailable, we use leave-one-out cross-validation to
find the model with the lowest mean squared prediction error.  Since
the exact conditionals from the alternative distribution are not
available in explicit form, we perform the cross-validation as follows:
\begin{enumerate}[(b)]
\item[(a)] Estimate $\Theta$ using all but one observation.
\item[(b)]
In the remaining\vadjust{\goodbreak} observation, estimate the values of the latent
  normal variables for all but one of the coordinates in the same manner as
  the variational E-step of Section \ref{sec:variational}.
\item[(c)] Predict the remaining normal value.
\item[(d)] Scale the
normal value by the expectation of $1/\sqrt{\tau}$.
\end{enumerate}
We remark that we also experimented with leaving out a larger fraction
of the observations as suggested in the work of \cite{shao}, but this
led to similar conclusions in the present example.

The cross-validation procedure gave a network with 122 edges.  To
reduce to the graph size found by \textit{GGM2}, we took 500
bootstrapped samples of the data, fixing the parameter $\rho$ found in
cross-validation, and only included those edges found in more than
$98.5\%$ of the samples.  For comparison, we also ran the above
procedure using the \textit{glasso}, but keeping $98\%$ of the samples
to obtain the same-sized graph.

\begin{figure}

\includegraphics{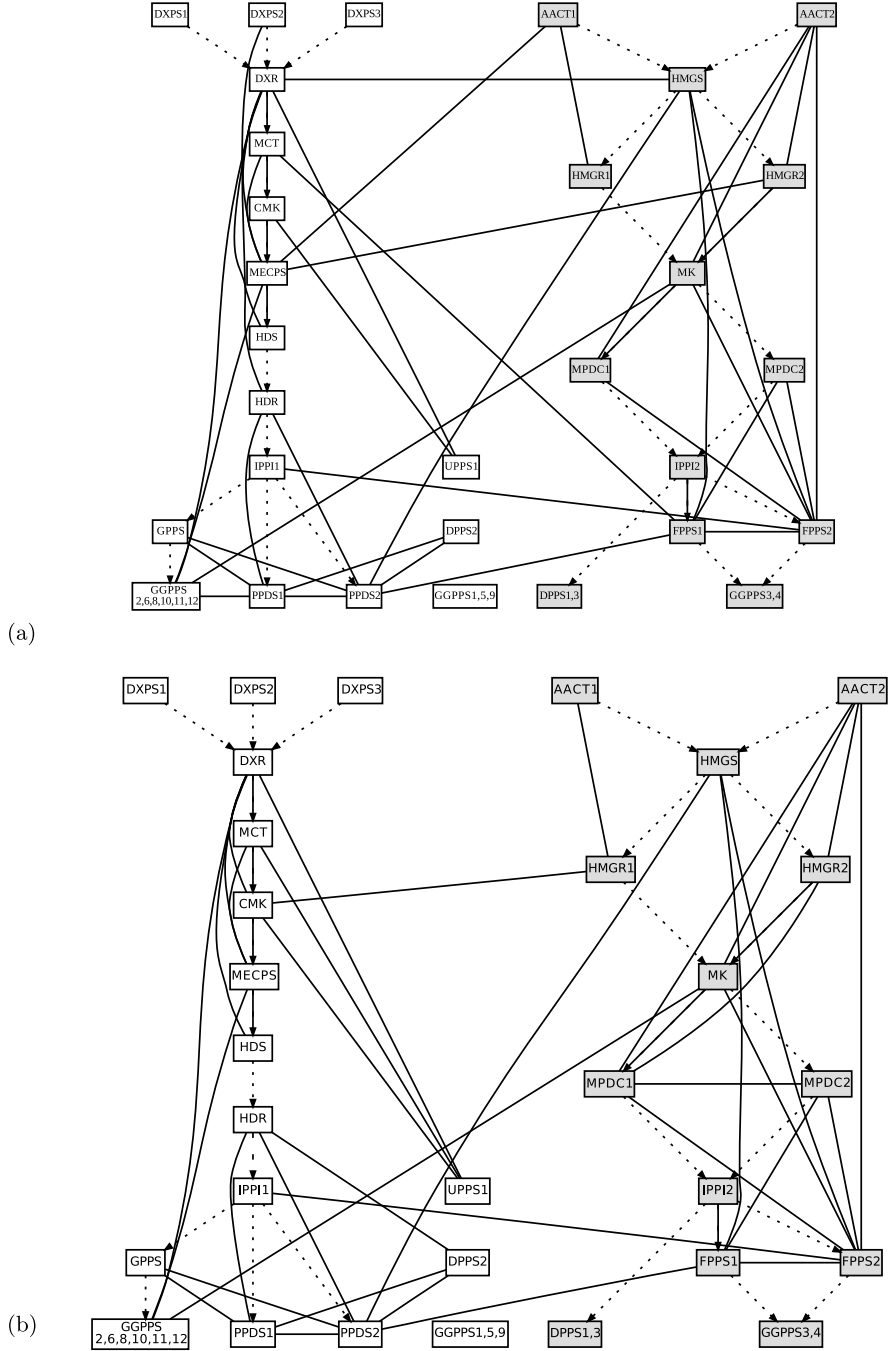}

\caption{Graphs recovered by  bootstrapping
procedure with target
  graph size of 43 using \textup{(a)} the \textit{glasso} and \textup{(b)} $t^*_{\mathrm{var}}$\textit{-lasso}.
  The graph shows the key relationships identified previously, but
  with fewer ``cross-talk'' edges.}\label{fig:isoprenoid2}
\end{figure}

We believe our procedure infers a graph that compares favorably (in
terms of biological interpretation) with that found by \textit{GGM2}.
Like \textit{GGM2}, we find a connection between AACT2 and the group
MK, MPDC1 and FPPS2; \textit{GGM1} found AACT2 to be disconnected from
the rest of the graph despite its high correlation with these three
genes. In the MEP pathway, our approach and \textit{GGM2} find similar
structure; compare Figures \ref{fig:isoprenoid1} and
\ref{fig:isoprenoid2}.

While our approach finds the key relationships identified in Wille
et al., it achieves this with fewer ``cross-talk'' edges between the two
pathways.  The authors discuss plausible interpretations for such
interactions between the pathways, but a graph with less cross-talk
might be closer to the scientists' original expectation
(Figures \ref{fig:isoprenoid1} and \ref{fig:isoprenoid2}). It is worth
noting that the \textit{glasso} procedure performs better than \textit{GGM1}, with edge inclusion being far less sensitive to the particular
bootstrapped sample.  The \textit{glasso} also finds the key
relationships of \textit{GGM2}.  We also ran the \textit{tlasso}, which
gave results similar to the \textit{glasso} and with the $t^*_{\mathrm{MC}}$\textit{-lasso},
 which behaved similar to the $t^*_{\mathrm{var}}$\textit{-lasso}.  We do
not show these results here.

\section{Discussion}\label{sec:discussion}

Our proposed \textit{tlasso} and $t^*_{\mathrm{var}}$\textit{-lasso}
algorithms are simple and effective methods for robust inference in
graphical models. Only slightly more computationally expensive than the
\textit{glasso}, they can offer great gains in statistical efficiency.
The \textit{alternative t} distribution is more flexible than the
classical $t$ and is generally preferred.  We find that the simple
variational E-step is an efficient way to estimate the graph in the
alternative case, but also explored more sophisticated Monte Carlo
approximations.

We assumed $\nu=3$ degrees of freedom in our various \textit{tlasso}
and $t^*$\textit{-lasso} runs.  As suggested in prior work on
$t$-distribution models, estimation of the degrees of freedom can be
done efficiently by a line search based on the observed log-likelihood
function in the classical model.

\begin{figure}

\includegraphics{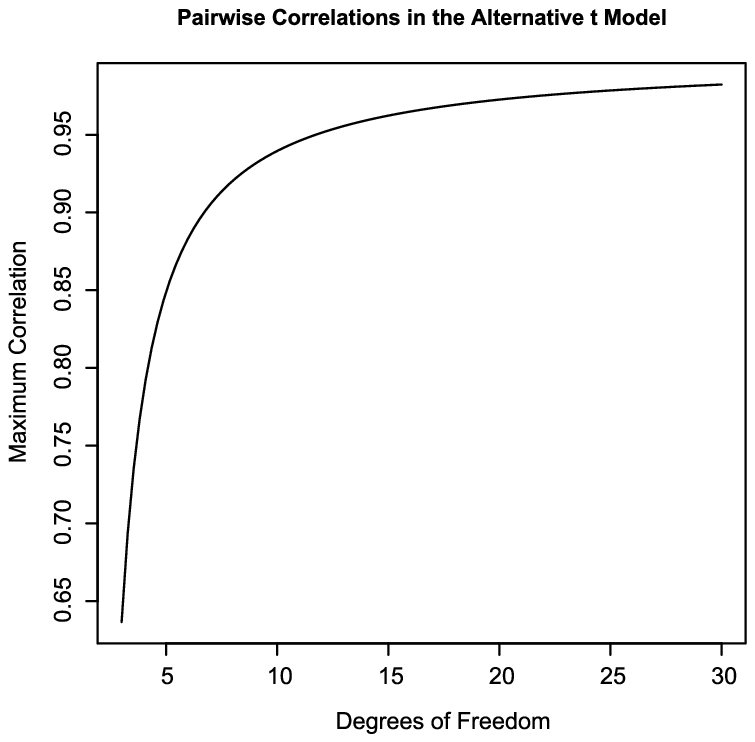}

\caption{The alternative $t$ model places an upper bound on the
correlation between two variables.  This bound increases with $\nu$,
but is fairly restrictive for the small degrees of freedom we
consider.} \label{fig:maxcorr}
\end{figure}

In the alternative model, the choice of $\nu$ puts an explicit upper
bound on the maximum correlation between two variables, the upper bound
increasing quickly with $\nu$ (see Figure \ref{fig:maxcorr}).  This
makes inference of the degrees of freedom potentially more relevant
than with the classical model, as an alternative model with small $\nu$
might not be a good fit for highly correlated variables.  In order to
select $\nu$, a line search based on the hidden log-likelihood function
can be employed.  For further flexibility, we may also allow the
degrees of freedom to vary with each node.  That is, we could let the
divisors $\tau_j \sim \Gamma(\nu_j/2,\nu_j/2)$ be independent
$\Gamma$-divisors with possible different degrees of freedom $\nu_j$.
This leads to similar conditionals in the Gibbs sampler and the
resulting procedure is thus no more complicated.  Nevertheless, for the
purposes of graph estimation, our experience and related literature
suggest that not much is lost by considering only a few small values
for the degrees of freedom.  For instance, running the $t^*_{\mathrm{var}}$\textit{-lasso} procedure in Section \ref{sec:isoprenoid-pathway}
using $\nu=5$ produces a very similar result with one additional
cross-talk edge.

In the last section we used cross-validation to choose the one-norm
tuning parameter $\rho$.  The likelihood is not available explicitly
for the $t^*$-distribution and so we cannot easily use information
criteria for the $t^*$\textit{-lasso}.  Cross-validation often tends to
pick more edges than is desirable, however, when the goal is inference
of the graph and not optimal prediction.  An interesting but
potentially difficult problem for future research would be to develop
rules for choosing $\rho$ that control an edge inclusion error rate;
compare \cite{banerjee}; \cite{meinshausen}.

Throughout the paper, we have penalized all the elements of $\Theta$.
One alternative is to remove the penalty from the diagonal elements of
$\Theta$, since we expect all these to be nonzero.  This leads to
smaller estimated partial correlations, and we found it to result in
less stable behavior of the \textit{tlasso} in the sense of the number
of edges decreasing rather suddenly as $\rho$ increases.

Finally, we remark that other normal scale-mixture models could be
treated in a~similar fashion as the $t$-distribution models we
considered in this paper.  However, the use of $t$-distributions is
particularly convenient in that it is rather robust to various types of
misspecification, involves only the choice of the degrees of freedom
parameters for the distribution of Gamma-divisors, and maintains good
efficiency when data are Gaussian.

\begin{appendix}
\section*{Appendix}\label{appendix}

\begin{pf*}{Proof of Proposition \ref{thm:1}}
  According to standard graphical model theory [\cite{lauritzen}], it
  suffices to show that $Y_j$ and $Y_k$ are conditionally uncorrelated
  given $Y_{V\setminus\{j,k\}}$.  Partition $V$ into $a=\{j,k\}$ and
  $b=V\setminus \{j,k\}$. For a given value of~$\tau$,
\[
(Y_a |Y_b, \tau) \sim N_2\bigl(\mu_a - \Theta_{a,a}^{-1} \Theta_{a,b} (Y_b-
 \mu_b), \Theta_{a,a}^{-1}/\tau \bigr)
\]
  and
 \[
 (Y_j|Y_{k\,\cup\,b},\tau) \sim N\bigl(\mu_i -
 \theta_{jj}^{-1}\Theta_{j,k\,\cup\,
  b} (Y_{k\,\cup\,b}- \mu_{k\,\cup\,b}), \theta_{jj}^{-1}/\tau\bigr).
 \]
  Since $\theta_{jk}=0$,
 \[
 \mathbb{E}[Y_j|Y_{k \,\cup\, b},\tau] = \mu_j - \theta_{jj}^{-1}\Theta_{j,b} (Y_{b}-\mu_{b})=\mathbb{E}[Y_j|Y_b,\tau]
\]
  for any value of $\tau$. Therefore,
 \[
    \mathbb{E}[Y_j|Y_{k \,\cup\,b}] =
    \mathbb{E}[\mathbb{E}[Y_j|Y_{k\,\cup\,
    b},\tau]|Y_{k\, \cup\, b}]  =
    \mathbb{E}[\mathbb{E}[Y_j|Y_b,\tau]|Y_b] = \mathbb{E}[Y_j|Y_b],
\]
  which implies that $Y_j$ and $Y_k$ are conditionally uncorrelated given $Y_b$.
\end{pf*}
\end{appendix}

\printaddresses

\end{document}